\documentstyle[12pt]{article}
\newcommand{\be}{\begin{equation}}
\newcommand{\ee}{\end{equation}}
\newcommand{\hm}{h^{-1} Mpc}
\newcommand{\rapj}{Astrophys.J.}
\newcommand{\rmnras}{Mon.Not.R.astr.Soc}
\newcommand{\lam}{\lambda}
\newcommand{\etal}{{\it et al.}}

\catcode`\@=11
\@addtoreset{equation}{section}
\catcode`@=12

\begin{document}
\baselineskip 16pt
\centerline{\LARGE The power spectrum}
\centerline{}
\centerline{\LARGE in a strongly inhomogeneous Universe}
\centerline{}

\centerline{\Large Francesco Sylos Labini$^{(1,2)}$ and
Luca Amendola $^{(3)}$}

\centerline{\footnotesize ($^1$) Dipartimento di Fisica, Universit\`a di
Bologna, Italy}

\centerline{\footnotesize ($^2$) Dipartimento di Fisica,
Universit\`a di Roma
``La Sapienza''}
\centerline{\footnotesize P.le A. Moro 2, I-00185 Roma, Italy.}

\centerline{\footnotesize ($^3$) Osservatorio Astronomico di Roma,
Viale del Parco Mellini 84, I-00136 Roma, Italy}

\begin{abstract}
\baselineskip 16pt
A crucial issue in cosmology is the determination of the fluctuation
power spectrum.
The standard picture of the matter clustering, the Cold Dark Matter model (and
its variant), assumes that,
on  scales smaller than a certain
``flattening scale'' $\lambda_f$, the power spectrum increases
with the scale, while on  much larger scales it  decreases
so to match the tiny fluctuations observed in the microwave
background. The standard picture also assumes that, once a statistically
homogeneous  sample
is reached, the power spectrum amplitude is fixed, and any
major variation should
be attributed  to luminosity segregation.
 However, the determination of  $\lambda_f$ and of the absolute
  amplitude,
 if any, is still matter
of debate. In particular,
  there is no consensus on whether the turnaround has
been detected or not, and on the actual importance of
the luminosity segregation effect.  We show that,
due to the finiteness of the sample
 the standard analysis of self-similar (fractal) distributions
 yields  a   turnaround
for scales close to the survey scale,  and
a systematic amplitude shift with the survey scale.
We point out that
both features, bending and scaling,
 are in agreement with recent determination of the
power spectrum,
in particular with the  CfA2 spectrum.
We remark that the standard power spectrum is not a well defined
statistical tool to characterize the galaxy distribution when a homogeneity
 scale has not been reached.
 In order to perform an analysis that does
 not imply any a priori assumption one should study
the PS of the density,
rather than the PS of the density contrast.
\end{abstract}
\vspace{.3in}
Subject headings: galaxies: clustering --
galaxies: distances and redshifts

\newpage

\section{Introduction}

Identifying the scale at which our Universe becomes homogeneous,
if any, is a crucial task of contemporary cosmology.  There's no
doubt that the assumption of homogeneity worked pretty well so far;
the theoretical expectations concerning the primordial
nucleosynthesis and the microwave background, for instance, are nicely
confirmed by most, if not all, observations, and it would be
an extremely difficult job to invent a radically alternative model
able to share
 the same level of experimental support.
 However, when we come
to consider the homogeneity of the luminous matter at the present, the
situation becomes much less clear.
(We refer the reader to Baryshev \etal 1994 for
a review of the recent experimental data and for a discussion of
some ideas on alternative cosmological models.)
 Essentially all the currently
elaborated models of galaxy formation
 assume large scale homogeneity and
predict that the galaxy
power spectrum (PS),
that is the power spectrum  of the density contrast,
decreases both toward small scales and toward large
scales, with a turnaround somewhere in the middle, at a scale $\lambda_f$
that can be taken as separating ``small'' from ``large'' scales.
This picture assumes the existence
of a homogeneity scale $\lambda_h>\lambda_f$,
defined as the scale at which the average density becomes
constant.
Then, because of the assumption of homogeneity, the power spectrum amplitude
should be independent of the survey scale, any residual
variation being attributed to luminosity bias (or to the
fact that the survey scale has not yet reached the homogeneity scale).
 Most variations on the theme, as the inclusion of
hot dark matter, or a cosmological constant, and so on,
simply push $\lambda_f$ somewhat to larger values, but
do not
qualitatively change the scenario. However, the crucial clue to this
picture, the firm determination of the
scale $\lambda_f$, is still missing, although
some surveys do indeed produce a  turnaround
 scale around 100 $\hm$
(Baugh \& Efstathiou 1992;
Feldman \etal 1994). Recently, the CfA2  survey
analyzed by  Park \etal (1994; PVGH) (and confirmed by SSRS2
- Da Costa \etal (1994, DVGHP)), showed a $n=-2$ slope up to $\sim 30 \hm$,
a milder $n\approx -1$ slope  up to 200 $\hm$, and some tentative
indication of flattening on even larger scales. They also find
that deeper subsamples have higher power amplitude,
i.e. that the amplitude scales with the sample depth.

In this paper we argue  that both features, bending and scaling,
are a manifestation of
 the finiteness of the survey volume, and that they
cannot be
interpreted as the convergence to homogeneity, nor to a power spectrum
flattening.
The systematic effect of the survey finite size is in fact
 to suppress
power at large scale, mimicking a real flattening.
Clearly, this effect occurs whenever galaxies have
a large correlation scale, with respect to the survey, and it has
often been studied in the context of standard scenarios (Itoh \etal 1992;
Colombi \etal 1994).
 We push this argument further, by showing that
even a fractal distribution of matter, i.e. a distribution which
never reaches  homogeneity, shows a sharp flattening. Such a
flattening is  partially corrected, but not quite eliminated,
  when the correction proposed by Peacock \& Nicholson (1991) is
applied to the data.
 We show also
 how the amplitude of the
power spectrum depends on the survey size as long as
the system shows long-range correlations.

The standard  power spectrum (hereafter SPS)
measures directly the contributions of different scales to the galaxy
density contrast $\:\delta\rho/\rho$.
It is clear that the density contrast, and all the quantities based on it,
is meaningful only when one can define
a constant density, i.e. reliably identify
the sample density with
the average density of all the Universe.
When this is not true, and we argue that is
indeed an incorrect assumption
in all the cases investigated so far, a false interpretation of the results may
occur, since both
the shape and the amplitude of the power spectrum depend on the
survey size.
Indeed, in the case of fractal structures the average
density  is not a
 well defined quantity, because it depends on the sample depth,
 as
Coleman \& Pietronero (1992, hereafter CP92) have shown for the
CfA1 redshift survey (see also Borgani 1995).
We stress  that a sample that contains
a portion of a fractal distribution is a statistically fair sample,
even being not homogeneous. While we focus here on the power spectrum
it is clear that in a general discussion of fractal versus homogeneity
one should consider many other observational effects. To this aim
we refer the reader to CP92, Baryshev et al., 1994, and
Sylos Labini et al., 1995.

In Sec.  2 we recall the basic
formulas of the SPS analysis (Peebles 1980),
 and we apply them to
the case of a fractal distribution.
Then we show that an analysis independent of the average density
is better suited to the description of the clustering when the
distribution is not homogeneous; in the case
of a fractal,
this scale-independent power spectrum
does not display any flattening, and its amplitude does not scale with
the sample size.
Of course it gives the same results of the SPS at the scales where
the distribution is homogeneous.
In Sec. 3 we
show that the results of the CfA2 analysis
  are in agreement with
our predictions in the case of a fractal distribution.
Our conclusion, sketched in the final section,  is that this analysis  supports
 the fractal nature of the CfA2 sample up to the survey scale, pushing
the spectrum turnaround,
and consequently the homogeneity scale, longward of 200 $\hm$.

\section{The  power spectrum of a fractal distribution}

Let us recall the basic  notation of the power spectrum analysis.
Following Peebles (1980)  we imagine that the Universe is periodic
in a volume $\:V_{u}$, with $\:V_{u}$ much
larger than the (assumed) maximum
correlation length. The survey volume $V\in V_u$
 contains $\:N$ galaxies at positions $\:\vec{r_i}$,
  and the galaxy density contrast is
$
\delta(\vec{r}) = [n(\vec{r})/\hat n] -1
$
where it is assumed that exists a
well defined constant density $\hat n$, obtained
averaging over a sufficiently large scale.
The density function
can be described by a sum of delta functions:
$~n(\vec{r}) = \sum_{i=1}^{N} \delta^{(3)} (\vec{r}-\vec{r_{i}})\,.~$
Expanding the density contrast   in its Fourier components we have
\be
\label{e7}
\delta_{\vec{k}} = \frac{1}{N} \sum_{j \epsilon V}
e^{i\vec{k}\vec{r_{j}}} - W(\vec{k})\,,
\ee
where
$~W(\vec{k}) = V^{-1} \int d{\vec{r}} W(\vec{r})
 e^{i\vec{k}\vec{r}}\,~$
is the Fourier transform of the survey window $W(\vec{r})$,
defined to be unity inside the survey region, and zero outside.
If $\xi(\vec{r})$ is the correlation function of the galaxies,
($\xi(\vec{r}) = <n(\vec{r})n(0)>/\hat n^2 -1$)
the true PS $\:P(\vec{k})$ is defined as
the Fourier conjugate of the
correlation function $\xi(r)$.
Because of isotropy the PS can be simplified to
\be
\label{e12}
P(k)  =4\pi \int \xi(r)  \frac{\sin(kr)}{kr} r^{2}dr\,.
\ee
The  variance of $\:\delta_{\vec{k}}$
is (Peebles 1980)
$<|\delta_{\vec{k}}|^{2}> = N^{-1}+ V^{-1}\tilde P(\vec{k})\,.
$
The first term is the usual additional shot noise term
while the second is the true PS convolved with a window function
that describe the geometry of the sample
\be
\label{e9b}
\tilde P(\vec{k})=
{V\over (2\pi)^3} \int
<|\delta_{\vec{k'}}|^{2}>
|W(\vec{k}-\vec{k'})|^2
d^3 \vec{k'}
\,.
\ee
%

We apply now this standard analysis to a fractal distribution.
In a self-similar system the number of points inside
a certain radius $\:r$  scales
according to the mass-length relation
(Mandelbrot 1982)
$~N(r) = Br^{D}\,,~$
with $\:D<3$ (the case $\:D=3$ corresponds to the
homogenous distribution) and the
constant
$\:B$ is related to the lower cut-offs.
The average density  for a spherical sample of radius
$\:R_s$ is therefore
$~\hat n = N(R_s)/V(R_s) = (3/4\pi) BR_s^{-(3-D)}\,.~$
It is simple to calculate the expression
of the $\:\xi(r)$ in this case (CP92)
\be
\label{e16}
\xi(r) = [(3-\gamma)/3](r/R_{s})^{-\gamma} -1\,,
\ee
 where  $\:\gamma=3-D$.
Notice that the volume integral of $\xi(r)$ over the sphere
 is bound to vanish,  since $\hat n$ is
 calculated from the data themselves. This implies ``anticorrelation''
 for $r$ close to $R_s$.
On scales larger that $R_s$ the $\xi(r)$ cannot be calculated without
making assumptions on the distribution outside the sampling volume
(in particular, assumption of homogeneity),
which is just what we want to avoid.
When the survey volume is not spherical, the scale $R_s$ is
 of the order of the largest sphere completely contained inside
the survey.
Since in a fractal  the average density depends on the scale, quantities like
$\xi(r)$
are also scale dependent.
{}From the Eq. (\ref{e16}) it follows  that for a fractal:
i)  the so-called correlation length
 $\:r_{0}$ (defined as $\:\xi(r_{0})= 1$)
is a linear function of the (spherical) sample size $\:R_{s}$:
$
r_{0} = [(3-\gamma)/6]^{\frac{1}{\gamma}}R_{s}.
$
A linear dependence of $\:r_{0}$ on the sample size $\:R_{s}$
has indeed been found in the whole CfA(1) sample (CP92);
ii) $\:\xi(r)$ is power law only for
$
[(3-\gamma)/3](r/R_{s})^{-\gamma}  \gg 1\,,
$
hence for $\: r \leq  r_{0}$; for larger
distances there is a
clear deviation from a power law behaviour.
Both
the amplitude and the shape of $\xi(r)$ are
therefore scale-dependent  in the
case of a fractal distribution.
It is clear  that the same kind of
finite size effects  are also present when computing the SPS.

The SPS for a model described by
Eq. (\ref{e16}) inside a sphere of radius $R_s$ is
\be
\label{e19}
P(k) =  \int^{R_{s}}_{0}
   4\pi \frac{\sin(kr)}{kr} \left[ \frac{3-\gamma}{3}
\left(\frac{r}{R_{s}}\right)^{-\gamma} -1\right] r^{2}dr=
\frac{a(k,R_s) R_{s}^{3-D}}{k^{D}}-\frac{b(k,R_s)}{k^{3}}\,.
\ee
Notice that the integral has to be evaluated inside $R_s$
because we want to compare $P(k)$  with its {\it estimation}
 in a finite size spherical survey of scale $R_s$.
 In the  general case, we must deconvolve the
 window contribution
 from $P(k)$; $R_s$ is then a characteristic window scale.
Eq. (\ref{e19})  shows the two scale-dependent features of the PS. First,
the amplitude of the PS
depends on the sample depth.
Secondly,
the shape of the PS
is characterized by two  scaling regimes:
the first one, at high wavenumbers,
is related to the fractal dimension of the
distribution in real space,
while the second one arises only because of
the finiteness of the sample.

In the case of $\:D=2$  in eq.\ref{e19} one has:
$~a = \frac{4\pi}{3} (2+\cos(kR_{s}))\,,~$ and
$~b = 4\pi \sin (kR_{s})\,.~$
The PS is then a power-law with exponent
$\:-2$ at high wavenumbers,
it flattens at low wavenumbers and reaches a maximum at
$k\approx 4.3/R_s$, i.e. at a scale $\lam\approx 1.45 R_s$.
The scale at which the transition occurs
is thus related to the sample depth.
In a real survey, things are complicated by the window function,
so that the flattening (and the turnaround) scale can only be determined
numerically.
We study this behaviour in detail in Sec.  3.

As we have seen,
the analysis of a distribution on scales at which homogeneity is not reached
must avoid the normalization through the mean density, if the goal is
to produce
results which are not related to the sample size, and thus misleading.
To this aim,  now we
consider the scale-independent PS (SIPS) of the density $\:\rho(\vec{r})$,
a quantity that  does not involve the computation
of the average density, and thus  gives
 an unambiguous information
of the statistical properties of the system.
We first introduce the density correlation function
$G(\vec{r}) = <\rho(\vec{x}+\vec{r})\rho(\vec{x})> = A r^{-(3-D)}\,,
$
where the last equality holds in the case
of a fractal distribution with dimension {\em D},
and where {\it A} is a constant
determined by the lower cut-offs of the distribution (CP92). Defining
the SIPS  as the Fourier conjugate of the correlation function $G(r)$,
one obtains that in a finite spherical volume
$\Pi(k) \sim A' k^{-D}$,
(where $A'=4\pi(1-\cos(k R_s))$ if $D=2$)
so that the SIPS is a single power law extending all over the system size,
without  amplitude  scaling
 with the sample size (except for $kR_s\ll 1$).
In analogy to the procedure above,
we consider the Fourier transform of the density
$\rho_{\vec{k}} = V^{-1} \sum_{j\in V} e^{-i\vec{k}\vec{x_j}}\,,
$
and its variance
$<|\rho_{\vec{k}}|^{2}> = V^{-1}\tilde \Pi(\vec{k})+ N^{-1} \,,
$
where $\tilde \Pi(\vec{k})$ is the same as in Eq. (\ref{e9b}),
with $<|\rho_{\vec{k'}}|^{2}>$ instead of $<|\delta_{\vec{k'}}|^{2}>$.

\section{Tests on artificial distributions}
 To study in detail the
finite size effects in the determinations of the PS
we have performed some tests on artificial distributions
with a priori assigned properties.
We distribute the sample  in a cubic volume $\:V_{u}$.
We determine $ P(k)$ ($ \Pi(k)$)
defined as the directionally averaged $\: P(\vec{k})$
 ($\: \Pi(\vec{k})$).
In practice the spatial covarage for any survey is
incomplete in solid angle and /or in depth; this
requires the introduction of  a window
function $\:W(\vec{x})$ that is unity inside the survey region and
vanishes outside. Following PVGH, the estimate of the
noise-subtracted PS given in Eq. (\ref{e9b})
for a strongly peaked window function is
\be
\label{e40}
 P(\vec{k}) = \left( <| \delta_{k}|^2> - \frac{1}{N} \right)
\left( \sum_{\vec{k}} |W_{\vec{k}}|^2 \right) ^{-1} (1-|W_{\vec{k}}|^2)^{-1}\,,
\ee
where $\hat \delta_k$ and $W(\vec{x})$ are defined in Sec. 2.
For the lowest wavenumbers the power spectrum estimator (\ref{e40}) is not
acceptable, because then the window filter flattens sensibly (see e.g. PVGH).
The factor $\:(1-|W_{\vec{k}}|^2)^{-1}$ has been introduced
by Peacock \& Nicholson (1991)
as an analytical correction to the erroneous identification
of the sample density with the population density. However, the correction
itself rests on
the assumption that the power spectrum is flat on very large scales,
which is just the feature we are testing for. PVGH actually
correct their results by comparing them to the power spectra of
$N$-body simulations; their conclusion is that the power spectrum correction
is a procedure reliable for wavelengths smaller than $\sim 200\hm$.
We show below that
all the
 features of CfA2, and most importantly the flattening and
 the amplitude scaling,
are easily accounted for by  a distribution which
is fractal up to the sample size.

We have generated  $D\approx 2$ fractal distributions with the random
$\:\beta$-model
algorithm (Benzi et al. 1984). Then we
have constructed artificial volume limited catalogs with roughly the same
geometry of the CfA2 survey.
We have computed the
quantity  $\: P(k)$ from Eq. (\ref{e40})
averaging over 50 random  observers (located on one of
the particles) for each  realization . In Fig. 1
it is shown the $\: P(k)$ {\it vs.} the scale
$2\pi/k$ with and without the correction factor,
for some different survey scales $R_s$,
together
with the angle-averaged window power spectrum $\:|W_k|^2$.
 The slope of the
PS at high wavenumbers is $\: \approx - D$ in
agreement with  Eq.\ref{e19}.
As anticipated, the flattening at low wavenumbers is here completely
 spurious,  i.e. it is
due to the finite volume effects on
the statistical analysis performed. In fact, comparing
the PS at the various sample scales, one can see that
the flattening of the PS occurs always near the boundary of the sample.
The turnaround scale roughly follows the relation
$\lambda=1.5 R_s$ found in Sec. 2.
Notice that the  PS starts
flattening
before the window spectrum flattens; as in PVGH, the change
in slope occurs for $|W_k|^2<0.2$, value that they assumed as preliminary
condition
for the estimator (\ref{e40}) to be valid.
The amplitude of the power spectrum scales according to eq.\ref{e19}
In Fig. 2 it is shown the behaviour of $\: \Pi(k)$
computed from Eq. (\ref{e40})
with $<|\rho_k|^2>$ in place of $<|\delta_k|^2>$
and without the Peacock-Nicholson correction:
as predicted in the previous section,
its amplitude does not scale with the sample size and it is
characterized by a single power law behaviour, up to  very large scales
(where the two terms in the factor $A'$
 of $\Pi(k)$ are comparable) .
Finally, in Fig. 3 we compare directly the PS of our artificial catalogs with
the  PS of the CfA2 subsamples
 obtained generating two volume limited
subsamples at 130 and 101 $\hm$ (PVGH).
The physical scale has been computed matching the CfA2-101 galaxy
average density.
  Both the shapes and the amplitudes
are  compatible with a fractal distribution.
As it can be seen,
for CfA2-101 the agreement is excellent; for CfA2-130 the two curves
are compatible inside the errors.

\section{Discussion and conclusions}

We have analyzed the PS of the galaxy density contrast  in
artificial
fractal distributions with the same
technique used for real redshift surveys (Peacock \& Nicholson 1991;
 Fisher et al. 1993; PVGH; DVGHP).
As we have shown, the standard analysis on a pure fractal would lead to the
 conclusions that the fractal distribution has a spectrum approximated  by two
power laws, that a turnaround occurs on the largest scales, and that
the amplitude scales with the sample depth; these conclusions are clearly
dependent on the size of the sample,
 and have nothing to do with the real
distribution.
The fact that recent evaluations of the galaxy PS showed just
these features motivated us to consider whether or not a pure fractal can
explain the
observational data. Let us remark that
the scaling effect is particularly important in evaluating
 the matter/galaxy bias factor.
Keeping the matter power spectrum fixed, the galaxy PS scaling
  as $\sim R_s^{3-D}$ implies indeed a
scaling of the bias factor $b\sim R_s^{(3-D)/2}$.

Let us summarize the results of
PVGH, by confronting them with
the analysis of the PS for
a fractal distribution:
i) for $\:k \ge 0.25$ ($\:\lambda \le 25 h^{-1}$ Mpc) the PS
in a volume limited sample
is very close to a power law with slope $\:n=-2.1$. In our view, this is the
behaviour at high wavenumbers
connected with the real fractal dimension.
ii) For  $\:0.05\le k \le 0.2$  ($120\hm>\lambda>30 \hm$) and the spectrum
is less steep, with a slope  about $\:-1.1$.
This bending is, in  our view, solely due to the
finite size of the sample.
iii) The  amplitude of the
volume limited subsample CfA2-130
PS is $\:\sim 40\%$ larger
than for CfA2-101.
This linear scaling of the amplitude can be
understood again considering that the sample is
fractal with $\:D=2$.
The last point deserves some more remarks.
It is worthful to notice
that this trend is qualitatively
confirmed by the results of Peacock \& Nicholson (1991),
who find a higher PS amplitude for a deep radio-galaxy survey,
and of Fisher \etal (1993) who find a lower amplitude
for the shallow IRAS 1.2Jy survey.
The same behaviour has been found in the analysis of the
$\:\xi(r)$:  $\:r_{0}$
scales linearly with the sample depth according to Eq.\ref{e16} (CP92).
The authors (PVGH)
explain this fact considering the dependence
of galaxy clustering on luminosity: brighter galaxies
correlate more than fainter ones. They support this
interpretation observing that brighter galaxies tend to avoid
underdense regions; they also analyze separately two subsets of
the same volume-limited sample of CfA2, one with the brighter half
of galaxies, the other with the fainter one, and  find in some cases
a correlation
between luminosity and  amplitude. It is certainly possible that
both mechanisms, the luminosity segregation and
the intrinsic self-similarity of the distribution,
are correct, and
each one explains part of the scaling.
However, PVGH do {\it not}
 detect such a luminosity segregation for the two largest
subsamples, CfA101 and CfA130, to which we are comparing our analysis
 here.
It seems therefore that the amplitude scaling at these scales can be entirely
attributed to the fractal scaling.
Another piece of evidence against the dominant role of
the luminosity segregation has been put forward in
 Baryshev \etal  (1994).
In this paper
two volume limited subsamples
with the same absolute magnitude limit, $\:M \geq -20$, but with different
depth,  CfA1-80 (limited at $\: v\leq 8000 {\rm km ~sec}^{-1}$)
and CfA2-130, have been compared.
If the hypothesis of luminosity segregation holds
then one should find that there is not
difference between the amplitude of the correlation function (and of the PS)
computed in these subsamples, as they contain galaxies with the same
average absolute magnitude. On the contrary one finds that
the amplitude scales as predicted by Eq. \ref{e16}.
Our conclusion is then that the fractal nature of
the galaxy distribution can explain, to the scales
surveyed so far, the shift of the amplitude
with sample depth of the PS and of $\xi(r)$.

Finally we stress that the fractal dimension of
the galaxy clustering rises from
 $\:D\sim1.4$ for CfA1 to $\:D\sim 2$ for CfA2,
in agreement with the result of other independent surveys:
Perseus-Pisces (Guzzo \etal 1992; Sylos Labini \etal 1995),
and  ESP
(Pietronero \& Sylos Labini 1994).

\vspace{.3cm}
\centerline{\bf Acknowledgments}
\vspace{.3cm}
We wish to thank L. Pietronero for very useful discussions.

\newpage

\baselineskip 24pt
\section*{Figure Captions}
\begin{itemize}
\item[Fig. 1]
  Top panel: Power spectrum for a fractal distribution with dimension
$\:D\approx 2$ {\it vs.} the scale
$2\pi/k$ without the correction factor (smaller symbols) and with the
 correction factor (larger symbols).
The error bars, shown for clarity only on one
sample, represent the scatter among
the different observers of the same fractal.
The three set of points refer, from top to bottom,
to artificial fractal samples of
75 (open triangles), 50 (filled squares), 32.5 (crosses)
and 25 (open sqaures) $\hm$.
The straight line shows the slope $D=2$.
 The flattening and the turnaround at
high wavenumbers is spurious. In the
bottom panel the three window functions are shown.


\item[Fig. 2] The scale-independent power spectrum for the same fractal and
the same four
scales as in Fig. 1. Now the spectrum is a single power law, up to the
very few largest
scales as expected. The reference line has a slop $D=2$.
 The amplitude of the spectrum is now constant inside the errors.

\item[Fig. 3] Comparison of
 power spectra of fractal distribution (triangles)
 with the CfA2 survey (squares).
 In the top panel, we plot the PS of the subsample CfA2-130
 (PVGH) along with the PS of our artificial fractal distribution
 (without the error bars for clarity).
 In the bottom panel, we plot CfA2-101 (PVGH) and a subsample of
 the same fractal as above, with a correspondingly scaled depth.

\end{itemize}


\end{document}